\documentstyle[aps,pre,eqsecnum,epsf]{revtex}

\newcommand{\e}{\mathrm{e}}
\newcommand{\E}{\varepsilon}
\newcommand{\be}{\begin{equation}}
\newcommand{\ee}{\end{equation}}
\newcommand{\bea}{\begin{eqnarray}}
\newcommand{\eea}{\end{eqnarray}}
\newcommand {\ind}[1]{\mathrm{#1}}

\def \O{{\cal O}}

\def \la{\langle}
\def \ra{\rangle}

\def \ts{ \textstyle }
\def \half{{\textstyle {1 \over 2}}}

\begin{document}

\title{Surface critical behavior of random systems at the ordinary 
transition}

\author{M. Shpot$~{^{a,c,*}}$, Z. Usatenko$~{^{a,b,\S}}$, Chin-Kun 
Hu$~{^{b,d}}$}

\address{\it 
$~{^{a}}$ Institute for Condensed Matter Physics of the National 
Academy of Sciences of Ukraine Lviv, 79011, Ukraine, 
E-mail:pylyp@icmp.lviv.ua\\ 
$~{^{b}}$ Institute of Physics Academia 
Sinica, Taipei, 11529, Taiwan \\ $~{^{c}}$ University of Essen, Essen, 
45117, Germany \\ 
$~{^{d}}$ Department of Physics, National Dong Hwa 
University, Hualien 97401, Taiwan}
\vspace{-0.1cm}
\maketitle
\begin{abstract}
We calculate the surface critical exponents of the ordinary transition
occuring in semi-infinite, quenched dilute Ising-like systems.
This is done by applying the field theoretic
approach directly in $d=3$ dimensions up to the two-loop approximation, as 
well as in $d=4-\E$ dimensions. At $d=4-\E$ we extend,
up to the next-to-leading order, the
previous first-order results of the 
$\sqrt{\varepsilon}$ expansion by Ohno and Okabe
[Phys. Rev. B {\bf 46}, 5917 (1992)].
In both cases the numerical estimates for 
surface exponents are computed using Pad\'e approximants extrapolating the 
perturbation theory expansions. The obtained results  
indicate that the critical behavior of semi-infinite systems with quenched 
bulk disorder is characterized by the new set of surface critical exponents.

\end{abstract}

\section{Introduction}\label{INTRO}
The critical behavior of quenched random systems undergoing 
continuous phase transitions is of great interest.
It is well known that the critical behavior of ideal
pure bulk substances may be changed by introducing disorder.
A prominent result in the theory of quenched disordered systems is the 
Harris criterion \cite{Harris74} which states that the 
presence of disorder is relevant for those pure systems which have a 
positive specific heat exponent $\alpha$.
Thus, in the class of $O(N)$ symmetric $N$-vector
models in $d$ space dimensions the Ising model is the one of primary
interest, having $\alpha(d)\ge 0$.
The marginal value, $\alpha =0$, corresponds to $d=2$.
In this case the small amount of disorder produces a marginal perturbation.
This results in logarithmic corrections to the power-law singularities
without modification of the critical exponents with respect to the pure system
(for reviews see \cite{Shalaev94,Dots95}). 

In three-dimensional disordered Ising systems a new set of modified critical exponents
appears. This was confirmed by renormalization group (RG) calculations
\cite{HL74,X75,Lub75,GL76}, experiments \cite{B...83,M...86}, and
Monte-Carlo simulations \cite{CS86,MLT86}.
After pioneering investigations using the Wilson's RG and $\E$ expansion
\cite{HL74,X75,Lub75,GL76,Sh77,JK77}, scaling-field method \cite{NR82} and 
the massive field theory in three dimensions \cite{Jug83,SSh81,Sh88,BSh92}, 
there has been a great renewed interest to the subject in the last few years 
\cite{ShAnSo97,FHY99,PS00,FHY00,PV00}. An interesting relation between the random Ising model and 
"$N$-colored" tethered membranes has been given in \cite{WK98}. Various 
aspects of physics of quenched disordered systems are the subject of recent 
extensive MC investigations \cite{B-P98,WD98,PRR99}.

The presence of boundaries, which are inevitable in real systems,
leads to the new, surface physics. General reviews on
surface critical phenomena are given in \cite{Bin83,Die86a,Die97}.
It is now well known that there are several surface universality classes 
defining the critical behavior in the vicinity of boundaries, at the 
temperatures close to the bulk critical point. Each surface universality 
class is defined both by the bulk universality class and specific 
properties of a given boundary.

What happens with the surface critical behavior, when the quenched disorder
is introduced? First, the defects may be localized only at the boundary.
In \cite{DN90a}, the relevance-irrelevance criteria of the Harris type
concerning the weak surface disorder have been worked out.
In the case of the ordinary transition (which corresponds to the free boundary
conditions), it has been demonstarated that for $d>2$ the weak surface randomness
is an irrelevant perturbation. The same conclusion follows from rigorous
arguments \cite{Die98} and Monte Carlo simulations \cite{PS98}.
There is a strong evidence that a small amount of quenched surface disorder
is irrelevant at the special transition as well \cite{DN90a,Die98,DS98}.

Second, the random quenched disorder may be introduced in the bulk,
say, of a semi-infinite system bounded by a plane surface.
This will, in general, shift the critical temperature of the bulk phase transition,
and drive the system to another, "random" fixed point.
According to the Harris criterion, the Ising systems are of main concern here.
The change of the universality class of the bulk will affect
the critical behavior of the bounding surface. Thus one will expect
the new surface critical exponents to appear. And, in view of the 
aforementioned properties of boundaries, these new surface exponents 
should emerge irrespective of the presence of the disorder just at the 
boundary itself.

This is a situation typical for $d>2$. The problem of a RG
calculation of "disorder-induced" surface critical exponents was
first addressed by Ohno and Okabe in 1992 \cite{OO92}
by using the $\E$ expansion about the upper critical dimension $d=4$.
 In the marginal case $d=2$, the boundary critical 
behavior of random systems has been studied very recently
in a series of papers \cite{ILSS98,LI00,PCBI00}.

In the present paper, we present an attempt to go beyond the approaches
and approximations employed in \cite{OO92} in calculating the
surface critical exponents of random semi-infinite Ising-like models.
Our main caclulations are performed directly at $d=3$ using
the massive field theory approach \cite{Massive,Par80,DS94,Sh97,DS98} 
in the two-loop approximation. Moreover,
we extend, up to the next-to leading order of the $\sqrt{\E}$ expansion, the
previous first-order results of Ohno and Okabe \cite{OO92}.
In both cases, numerical estimates for surface
critical exponents of the ordinary transition
are evaluated using Pad\'e approximants improving the direct 
perturbation theory expansions. The obtained values of the surface critical 
exponents are consistent with the results by Ohno and 
Okabe \cite{OO92} and confirm that the semi-infinite systems 
with quenched bulk disorder are characterized by a new set of the surface 
critical exponents.

\section{Model}
The description of the surface critical behavior
at the ordinary transition can be formulated in terms of the
effective Landau-Ginzburg-Wilson Hamiltonian
\be\label{eh0}
H[\varphi]=\!\!\int_0^\infty\!\!\!\! dz\!\!\int\!\! d^{d-1}r \left [
\half |\nabla\varphi |^2 +\half \tau_0 \varphi^2 + \ts{1\over 
4!}v_0\varphi^4 \right ]\,. \ee
The $d$-dimensional spatial integration is extended over a half-space
$I\!\!R^d_+\equiv\{x{=}(r,z)\in I\!\!R^d\mid r\in I\!\!R^{d-1}, z\ge 0\}$
bounded by a plane surface at $z=0$.
Here $\varphi=\varphi(x)$ is a continuous scalar field corresponding to the
one-component order parameter of an original Ising system.
The surface is considered to be free, and hence the field $\varphi(x)$
satisfies Dirichlet boundary conditions at $z=0$,
$\varphi(r,0)=0$ \cite{DD81c,Die86a}.

One of the possibilities to introduce disorder into the model is
to assume that the parameter $\tau_0$ incorporates local random temperature
fluctuations $\delta\tau(x)$ via $\tau_0=m_0^2+\delta\tau(x)$.
Here $m_0^2$ is a "bare mass" representing linear temperature deviations
from the mean-field ctitical temperature, and the random variable
$\delta\tau(x)$ has the properties $\langle\delta\tau(x)\rangle_{\ind{conf}}=0$
and $\langle\delta\tau(x)\delta\tau(x')\rangle_{\ind{conf}}=\Delta\delta(x-x')$
with $\Delta >0$.
Angular brackets with the subscript {\it conf}
denote configurational averaging over quenched disorder which should be 
implemented on the level of the free energy \cite{Br59,GL68}.

Ohno and Okabe \cite{OO92} analyzed the above random model
by direct averaging over disorder using the method originally introduced by
Lubensky \cite{Lub75}.
Alternatively, the configurational averaging of the free energy can be performed
using the replica trick
\be\label{repl}
F=-T \lim_{n\to 0} {1\over n} \big (\la Z^n \ra_{\ind{conf}}-1\big),
\ee
where $Z$ is the partition function of a configuration given by the Boltzman 
weight $\e^{-H[\phi]}$,
as it was first done in the RG calculations by Grinstein and Luther \cite{GL76}.
We shall use this last possibility to treat randomness.
Employing the replica trick leads to the effective LGW Hamiltonian
with cubic anisotropy defined in the semi-infinite space,
\bea\label{eh}\nonumber
H[\vec\varphi]&=&\!\!\int_0^\infty\!\!\!\! dz\!\!\int\!\! d^{d-1}r \Big [
\half |\nabla\vec\varphi |^2 +\half m_0^2 |\vec\varphi |^2\\
 &+&\ts{1\over 4!}v_0\sum\limits_{\alpha=1}^n \varphi_{\alpha}^4 +
\ts{1\over 4!}u_0(|\vec\varphi|^2)^2 \Big ],
\eea
where $\vec\varphi (x)$ is an $n$-vector field with the components
$\varphi_i (x)$, $i=1,\cdots ,n$.
This last effective Hamiltonian is appropriate for the description of
the ordinary transition of random systems in the replica limit $n\to 0$.
The $\O(n)$ symmetric term arises due to
the random averaging in (\ref{repl}) via cumulant expansion and thus
its coupling constant $u_0\propto -\Delta < 0$.

\section{Correlation functions and their renormalizations}
The model defined in (\ref{eh}) is translationally invariant in directions
parallel to the boundary. Thus it is often useful to perform Fourier transformations
in $d-1$ dimensional subspace with respect to "parallel" coordinates $r$.
We shall denote the associated parallel momenta as $p$.
In the perpendicular direction the coordinate $z$ is retained.
Sometimes, in the perturbative calculations it is advantageous
to work in the complete coordinate representation, without any
transformations to the momentum space.
This is another difference with respect to the approach of
Ohno and Okabe \cite{OO92} who employ the full momentum-space representation
in all $d$ directions as it was originally done in the earliest
field-theoretical work on semi-infinite systems \cite{DD81c,RG81}.

The fundamental two-point correlation function of the free theory
corresponding to (\ref{eh}) is given by the Dirichlet propagator:
\be
\la \varphi_i(r,z)\varphi_j(0,z')\ra _0 = G_{\ind{D}}(r;z,z') 
\delta_{ij}\,. \ee
In the familiar $pz$ representation the Dirichlet propagator reads
\be\label{prop}
G_{\ind{D}}(p;z,z')={1 \over2\kappa_0}\left[e^{-\kappa_0|z-z^\prime|}-
e^{-\kappa_0(z+z^\prime)}\right],
\ee
where the standard notation is used, $\kappa_0=\sqrt{p^2+m_0^2}$.

The Dirichlet propagator vanishes identically when at least one of its $z$
coordinates is zero. Consequently, all the correlation functions involving
at least one field at the surface vanish. This property holds for both the
free and renormalized theories \cite{Die86a}.
Owing to this property, it is not a straightforward matter to analyze the
surface critical behavior at the ordinary transition.

In fact, the critical surface singularities at the ordinary transition
can be extracted by studying the correlation functions with
insertions of ({\it inner}) normal derivatives of the fields at the boundary,
$\partial_n\phi(r)$ \cite{DD80,DD81c,DDE83}. Actually, in order to obtain
the characteristic exponent $\eta_\parallel^{ord}$ of surface correlations,
it is sufficient to consider a correlation function of two
normal derivatives of boundary fields
\begin{equation}
{\cal G}_2(p)=\left\langle {\partial\over\partial 
z}\left.\varphi(p, z)\right|_{z=0} {\partial\over\partial 
z'}\left.\varphi(-p, z')\right|_{z'=0} \right\rangle,
\end{equation}
where the fields $\varphi(p,z)$ are the Fourier transforms of the fields 
$\varphi(r,z)$ in $d-1$ dimensional parallel subspace.
${\cal G}_2(p)$ is a parallel Fourier transform of the corresponding 
two-point function ${\cal G}_2(r)$ in direct space. At the critical point 
${\cal G}_2(p)$ behaves as $p^{-1+\eta_\parallel^{ord}}$.
It reproduces the leading critical behavior of a two-point function
$G_2(p)=\langle\varphi(p, z)\varphi(-p, z^{'})\rangle$ in the vicinity of 
the boundary plane. The surface critical exponent $\eta_\parallel^{ord}$ is 
given by the scaling dimension of the boundary operator 
$\partial_n\varphi(r)$.

In the presence of randomness, the exponent
$\eta_\parallel^{ord}$ differs from its counterpart in ordered
semi-infinite system. The other surface exponents of the ordinary transition
can be determined through the scaling laws \cite{Die86a}.

In the present formulation of the problem, the
renormalization process for the random system is essentially the same as in the
"pure" case \cite{Die86a,DS98}.
One introduces the renormalized bulk field and its normal derivative at the surface
through
\begin{equation}
\!\varphi_{R}(x){=}Z_\varphi^{-{1\over 2}}\,\varphi(x) \quad\mbox{and}\quad
\big(\partial_n\varphi(r)\big)_R{=}Z_{\parallel}^{-{1\over 
2}}\partial_n\varphi(r),\nonumber \end{equation}
and renormalized correlation functions involving $N$ bulk fields and $M$
normal derivatives,
\begin{equation}
{\cal G}_R^{(N,M)}(p;m,u,v)
=Z_\varphi ^{-{N\over 2}}Z_{\parallel} ^{-{M\over 2}} {\cal 
G}^{(N,M)}(p;m_{0},u_0,v_0) 
\end{equation}
for $(N,M)\ne (0,2)$. In the special case of only two surface
operators $(N,M)=(0,2)$, an additional, {\it additive} renormalization is 
required, so that 
\begin{equation}\label{addren}
{\cal G}_R^{(0,2)}(p)=Z_{\parallel}^{-1} \left[{\cal G}^{(0,2)}(p)-
{\cal G}^{(0,2)}(p{=}0)\right]\,.
\end{equation}
Everywhere the renormalizations of the mass and coupling constants are implicit.
These are the standard ones of the massive
infinite-volume theory.
The surface renormalization factor $Z_{\parallel}(u,v)$ can be
conveniently obtained from the consideration
of the boundary two-point function ${\cal G}^{(0,2)}$,
which is the object with the simplest Feynman graph expansion (see below):
\begin{equation}\label{zn}
Z_{\parallel}=-\lim_{p\to 0}{m\over p}{\partial\over\partial p}\,{\cal 
G}^{(0,2)}(p)\,. \end{equation}
A standard RG argument involving an inhomogeneous Callan-Symanzik equation
yields the anomalous dimension of the operator $\partial_n\varphi(r)$
\begin{eqnarray}\label{etan}
\eta_{\parallel}&=&m\!{\partial\over \partial m} 
\left.\ln\!{Z_{\parallel}}\,\right|_{F\!P}\\
      &=&\beta_u(u,v){\partial\ln Z_{\parallel}(u,v)\over \partial u}+
 \left.\beta_v(u,v){\partial\ln Z_{\parallel}(u,v)\over \partial v}\right 
 |_{F\!P}\,.\nonumber
\end{eqnarray}
"FP" indicates here that the above value should be calculated at the
infrared-stable random fixed point of the underlying bulk theory, 
$(u,v)=(u^*,v^*)$. The surface critical exponent $\eta_{\parallel}^{ord}$ at 
the ordinary transition is then given by 
\begin{equation}
\eta_{\parallel}^{ord}=2+\eta_{\parallel}\,.
\end{equation}

\section{Two-loop approximation}
The Feynman diagram expansion of the unrenormalized correlation function
${\cal G}^{(0,2)}$ is, to the two-loop order,
\begin{eqnarray}\label{23}
&&{\cal G}^{(0,2)}(p;m_0,u_0,v_0)=-\kappa_0+\epsfxsize=1.1cm 
\epsfbox{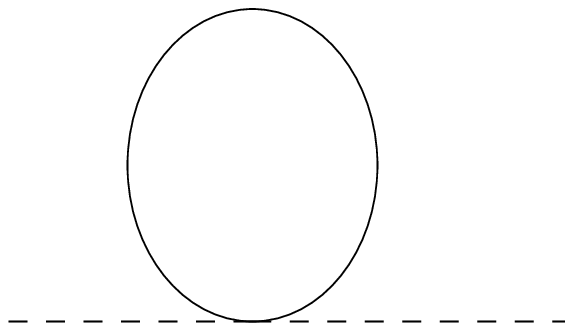}\nonumber\\ 
&&+\raisebox{-5pt}{\epsfxsize=1.7cm 
\epsfbox{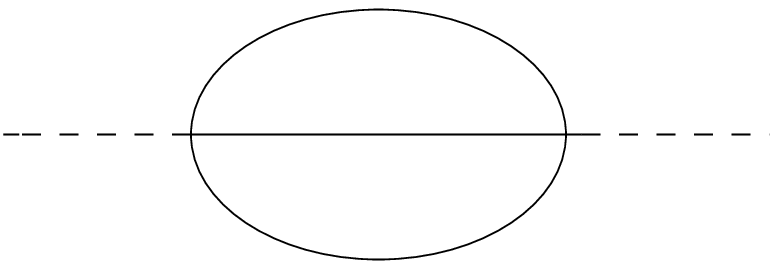}}+ \epsfxsize=2cm \epsfbox{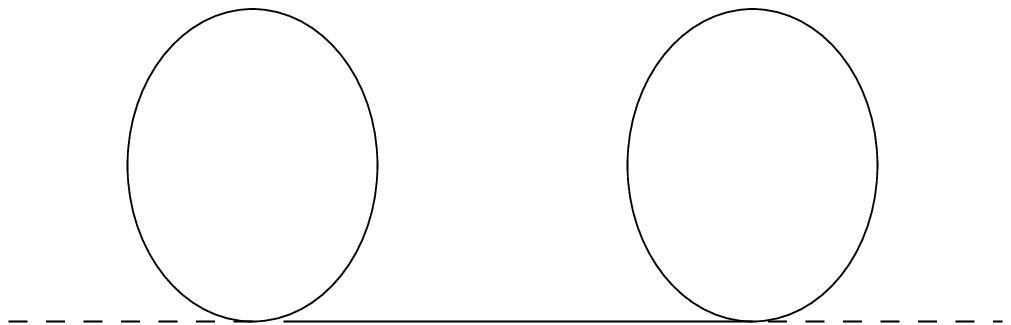}+
\epsfxsize=1.1cm \epsfbox{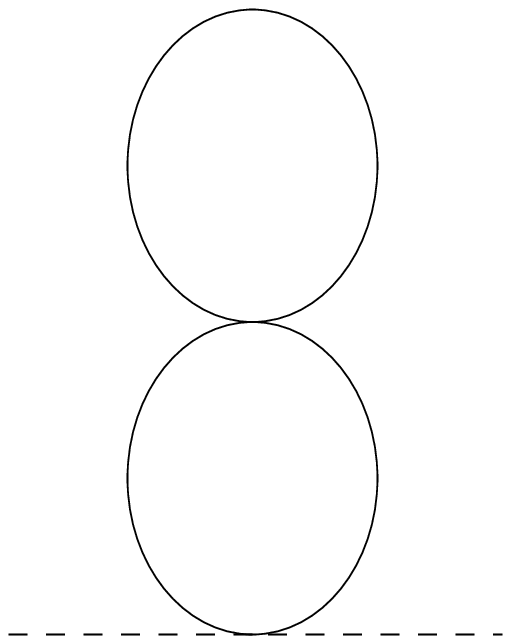}.
\end{eqnarray}
Full internal lines represent here the free Dirichlet propagators (\ref{prop})
and dashed external lines give the factors $e^{-\kappa_0 z_i}$
when attached to the internal point with the coordinate $z_i$.
Let us enumerate diagrams in the above sequence 1,2,3,4.
In the present theory described by the effective Hamiltonian (\ref{eh}),
these graphs have their corresponding weights 
\begin{mathletters}\label{ti}
\begin{eqnarray}
&-&{t_1^{(0)}\over 2}\quad\mbox{with}\quad t_1^{(0)}=\frac{n+2}{3}\,u_0+ v_{0}\,,\\
&&{t_2^{(0)}\over 6}\quad\mbox{with}\quad 
t_2^{(0)}=\frac{n+2}{3}\, u_{0}^{2}+ v_{0}^{2}+ 2v_{0}u_{0}\,,\\
&&{t_3^{(0)}\over 4}\quad\mbox{and}\quad {t_4^{(0)}\over 
4}\quad\mbox{with}\quad t_3^{(0)}=t_4^{(0)}=\big(t_1^{(0)}\big)^2.
\end{eqnarray}
\end{mathletters}

The typical bulk short-distance singularities, which are present in the 
graphic expansion (\ref{23}), are subtracted after performing the mass 
renormalization 
\begin{equation}
m_{0}^{2}= m^{2} + \epsfxsize=0.5cm \epsfbox{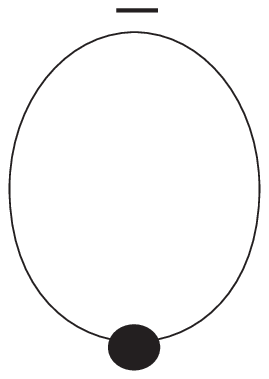}+\epsfxsize=0.9cm 
\epsfbox{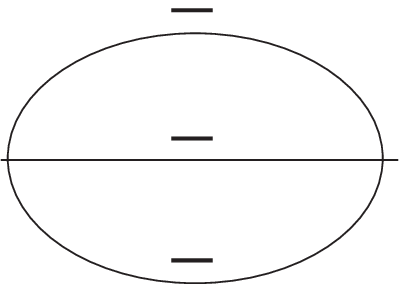} - m^2 \frac{\partial}{\partial k^2}\left. 
\epsfxsize=1cm \epsfbox{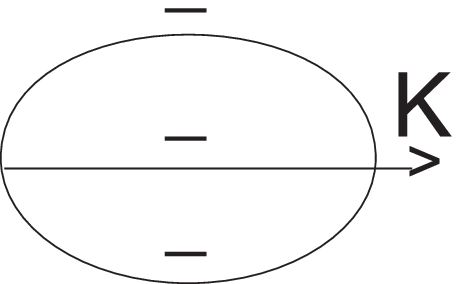}\right|_{k^2=0}.
\end{equation}

Here the full lines with sighns "-" denote the free {\it bulk} propagators.
They are associated with the {\it first} term in the 
Dirichlet propagator (\ref{prop}), which is the usual {\it bulk} massive 
propagator in the $pz$ representation. As a result of the mass 
renormalization one obtains 
\begin{eqnarray}\label{mr}
&&{\cal G}^{(0,2)}(p;m,u_0,v_0)=-\kappa-\epsfxsize=1.1cm 
\epsfbox{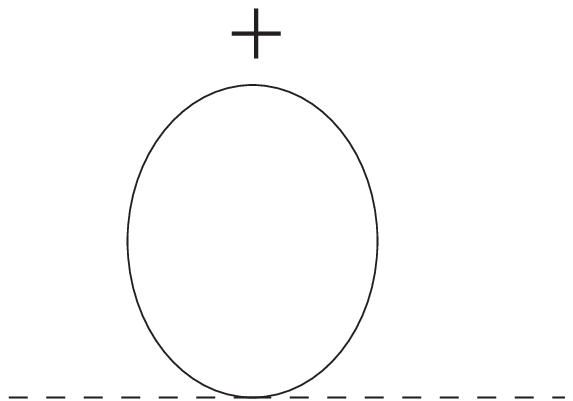}\nonumber\\ 
&&+\raisebox{-5pt}{\epsfxsize=1.7cm \epsfbox{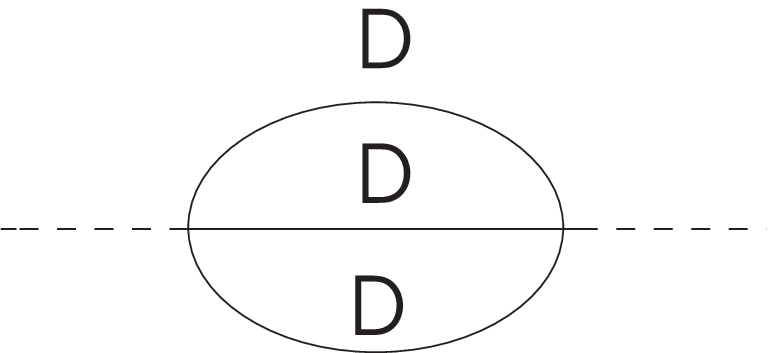}}
-\frac{1}{2\kappa}\raisebox{-5pt}{\epsfxsize=0.9cm \epsfbox{fig2b.eps}}+
\frac{m^2}{2\kappa}\frac{\partial}{\partial k^2}\left. 
\raisebox{-5pt}{\epsfxsize=1cm 
\epsfbox{fig2k.eps}}\right|_{k^2=0}\nonumber\\ 
&&+\epsfxsize=2.1cm 
\epsfbox{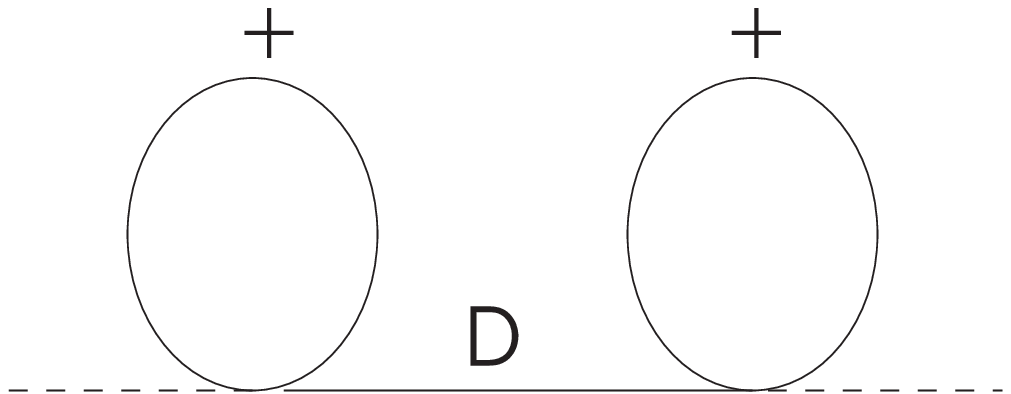}-\epsfxsize=1.1cm \epsfbox{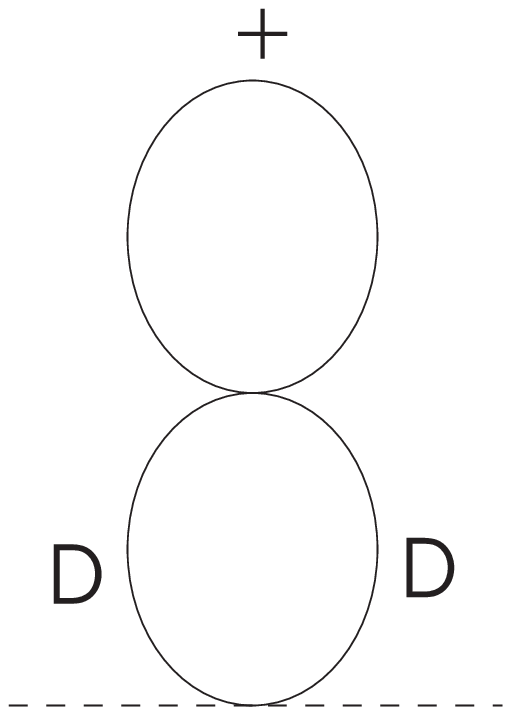}\;. 
\end{eqnarray} 

In this expansion the "surface" divergences still remain present, due to the 
one-loop self-energy insertions. These are represented by the closed lines 
with the index "+" which means only the {\it second}, mirror term of the 
free propagator (\ref{prop}). Moreover, we have explicitly indexed here, 
with the label "D", the original Dirichlet lines (cf. (\ref{23})). Bulk 
propagators are denoted with "-" sighns. The last two graphs in the second 
line represent just the usual bulk subtractions. 

The boundary singularity in the one-loop diagram of the two-point surface 
correlation function has to be removed through the additive 
renormalization [zero-momentum subtraction (\ref{addren})]. This, however, 
does not influence the calculation of the renormalization factor 
$Z_{\parallel}$ which involves a momentum differentiation (see 
(\ref{zn})). Surface divergences, present in each of the two last bubble 
graphs, mutually cancel. Actually, the whole combination in the third line 
of (\ref{mr}) vanishes identically as in \cite{DS98}. Hence, applying the 
rule (\ref{zn}), we obtain 
\begin{eqnarray}
&&Z_{\parallel}=1+{\partial\over \partial\kappa}
\left.\epsfxsize=1.1cm \epsfbox{fig11m.eps}\right |_{\kappa=m}
-\lim_{p\to 0}{m\over p}{\partial\over\partial p}\,
\raisebox{-5pt}{\epsfxsize=1.7cm \epsfbox{fig22m.eps}}
\nonumber\\
&&-{1\over 2 m^2}\left[\raisebox{-5pt}{\epsfxsize=0.9cm\epsfbox{fig2b.eps}} 
-m^2{\partial\over \partial k^2}\left. \raisebox{-5pt}{\epsfxsize=1cm 
\epsfbox{fig2k.eps}}\right|_{k^2=0}\right]\,.\label{znn} 
\end{eqnarray}
Performing the integration of (\ref{znn}) by analogy with \cite{DS98} we 
derive the result 
\begin{equation}
Z_{\parallel}(\bar u_0,\bar v_0)=1+{\bar t_1^{(0)}\over 4} +\bar 
t_2^{(0)} C,
\end{equation}
where the constant $C$ stems from the two-loop contribution into (\ref{znn}),
\begin{equation}\label{co}
C\simeq{107\over 162}-  {7\over 3} \ln{4\over 3}-0.094299\simeq -0.105063 \;.
\end{equation}
 
Here the renormalization factor $Z_{\parallel}$ is expressed as a 
second-order series expansion in powers of {\it bare} dimensionless 
parameters 
$\bar u_0=u_0/(8\pi m)$ and $\bar v_0=v_0/(8\pi m)$.
The corresponding weighting factors $\bar t_1^{(0)}$ and $\bar t_2^{(0)}$
are obtained by replacements
$(u_0,v_0)\to (\bar u_0,\bar v_0)$ in the original combinations
$t_1^{(0)}$ and $t_2^{(0)}$ from (\ref{ti}). As it is usual in {\it 
super}renormalizable theories, the renormalization factor expressed in terms 
of unrenormalized coupling constants is finite.

As a next step, the vertex renormalizations should be carried out.
To the present accuracy, they are given by
\begin{mathletters}
\begin{eqnarray}
\bar{u}_{0}&=&\bar{u}\Big(1+\frac{n+8}{6}\,\bar{u}+\bar{v}\Big),\\
\bar{v}_{0}&=&\bar{v}\Big(1+\frac{3}{2}\,\bar{v}+2\, \bar{u}\Big)\,.\label{33}
\end{eqnarray}
\end{mathletters}
Again, the vertex renormalization at $d=3$ is a finite reparametrization.
All relevant singularities have been removed already after the mass
renormalization and taking into account the special bubble-graph combinations
emerging in the theory with {\it Dirichlet} propagators.
The result is a modified series expansion 

\begin{eqnarray}
Z_{\parallel}(\bar u, \bar v)&=&
1+{n{+}2\over 12}\,\bar u +{\bar v\over 4}
+{n{+}2\over 3}\left(C+{n{+}8\over 24}\right)\,\bar u^2\nonumber\\
&+&\left(C+{3\over 8}\right)\,\bar v^2+2\left(C+{n{+}8\over 24}\right)\,\bar u\bar v\,.
\end{eqnarray}

Combining the renormalization factor $Z_{\parallel}(\bar u, \bar v)$ 
together with the one-loop pieces of the beta functions
\begin{mathletters}
\begin{eqnarray}
\beta_{\bar u}(\bar u, \bar v)&=&-\bar u\Big(1-{n{+}8\over 6}\;\bar u-\bar 
v\Big),\\ 
\beta_{\bar v}(\bar u, \bar v)&=&-\bar v\Big(1-{3\over 2}\,\bar 
v-2\bar u\Big), \end{eqnarray}
\end{mathletters}
through (\ref{etan}) yields the desired series expansion for 
$\eta_{\parallel}$. 

In terms of renormalized coupling constants $u$ and $v$, normalized in a 
standard fashion so that $u{=}{n+8\over 6}{\bar{u}}$ and $v{=}{3\over 
2}{\bar{v}}$, we obtain finally 
\begin{eqnarray}\label{etafin}
&&\eta_{\parallel}(u,v)=-{n{+}2\over {2 (n{+}8)}}\,u-{v\over 6}\\
&&-24{(n{+}2)\over (n{+}8)^2}{\cal C}(n) u^2-{8\over 9}{\cal 
C}(1) v^2 -{16\over n{+}8}{\cal C}(n) uv ,\nonumber
\end{eqnarray}
where ${\cal C}(n)$ is a function of the replica number $n$, defined as 
\begin{equation}
{\cal C}(n)=C+{n{+}14\over 96}.
\end{equation}

In fact, the last expression (\ref{etafin}) for $\eta_{\parallel}$ 
provides a result for the {\it cubic anisotropic} model given by the effective 
Hamiltonian (\ref{eh}) with general number $n$ of order-parameter components.
In the case of infinite space, this last model attracted much 
attention very recently (see e.g. \cite{PS00,Var00,CPV00} and references 
therein).

In the following we restrict our discussion to the case of {\it random Ising} system
by taking the replica limit $n\to 0$. Hence, we obtain the next two-loop 
expansion for the surface critical exponent $\eta_{\parallel}^{ord}$
\begin{equation}\label{epars}
\!\!\eta_\|^{ord} =2-{u\over 8}-{v\over 6}-{3\over 4}{\cal C}(0)u^2-
{8\over 9}{\cal C}(1)v^2-2{\cal C}(0)uv.
\end{equation}
Through the scaling relations we get access to the other surface critical 
exponents. For convenience further below we suppress the superscript {\it 
ord} at the surface critical exponents. 

\section{Surface critical exponents}

As it is well-known, the knowledge of one certain exponent
at the ordinary transition allows one to define
the complete set of other surface critical
exponents through the scaling relations \cite{Die86a}. For convenience we 
quote them here: 
\begin{eqnarray}\label{sc}
&& \eta_{\perp} = \frac{\eta + \eta_{\parallel}}{2}, \nonumber\\
&& \beta_{1} = \frac{\nu}{2} (d-2+\eta_{\parallel}), \nonumber\\
&& \gamma_{11}=\nu(1-\eta_{\parallel}), \nonumber\\
&& \gamma_{1}= \nu(2-\eta_{\perp}), \nonumber\\
&& \Delta_{1}= \frac{\nu}{2} (d-\eta_{\parallel}), \nonumber\\
&& \delta_{1} = \frac{\Delta}{\beta_{1}} = 
\frac{d+2-\eta}{d-2+\eta_{\parallel}}, \nonumber\\
&& \delta_{11} = \frac{\Delta_{1}}{\beta_{1}}=
\frac{d-\eta_{\parallel}}{d-2+\eta_{\parallel}}\;.\nonumber
\end{eqnarray}

The exponent $\eta_\perp$ characterizes the critical-point correlations
perpendicular to the surface, $\beta_1$ describes the decay of the surface 
magnetization on approaching the critical temperature,
$\gamma_{11}$ is the (local) surface  susceptibility exponent,
$\gamma_1$ is the layer susceptibility exponent,
$\Delta_1$ is the surface magnetic shift exponent, and
$\delta_{11}$ and $\delta_1$
give relations between the surface magnetization and the
surface and bulk external magnetic fields, respectively,
along the critical isotherm.
The values $\nu$, $\eta$, and $\Delta=\nu(d+2-\eta)/2$
are the standart bulk exponents.

In order to obtain individual RG expansions for each surface exponent,
we use the above scaling laws (with $d=3$) combining
Eq. (\ref{epars}) with the $n\to 0$ limits of the
two-loop series expansions for bulk
exponents $\nu$ and $\eta$ \cite{SSh81,Jug83,Sh88a}

\begin{eqnarray}
&& \nu=\frac{1}{2} \left[1+ \frac{v}{6} + \frac{(n+2)}{2(n+8)} u 
\right.\nonumber\\
&& \left.-\frac{1}{324} \left[ \frac{11}{9} v^{2}- \frac{2}{n+8}(27n-38) 
u v - \frac{3 (n+2)}{(n+8)^2} (27n-38) u^{2} \right]\right],\\
&& \eta=\frac{8}{27} \left[ \frac{ v^{2}}{27} + 
\frac{2 u v}{3(n+8)} + \frac{(n+2)}{(n+8)^2} u^{2}\right]. \label{46}
\end{eqnarray}

For each of surface critical exponents we obtain a double series expansion 
in powers of $u$ and $v$ of the form 
\begin{equation}
f(u, v) = \sum_{j,l= 0}^{\infty} f_{jl} 
u^{j} v^{l}, \label{48}
\end{equation}
truncated at the second order. Power series expansions of this kind 
are known to be generally divergent
due to a nearly factorial growth of expansion coefficients at large orders
of perturbation theory \cite{LO,lo87,McK94,AMR00}. Hence, the
numerical evaluation of the exponents represented by such series expansions requires
additional "resummation" procedures.

The simplest way to obtain meaningful and rather accurate numerical
estimates is to construct a table of rational approximants in two 
variables from the original series expansions. 
This should work already well when the series behave in lowest orders
"in a convergent fashion". Apparently divergent ones require more sophisticated
summation procedures.

The results of our Pad\'e analysis are represented in Table I.
We evaluate the exponents at the standard two-loop random  
fixed point {\cite{Jug83}}
\begin{equation}\label{NT}
u^{*}=-0.60509, \quad\quad\quad v^{*}=2.39631.
\end{equation}

To give an idea about relative magnitudes of first- ($O_1$) and second-order
($O_2$) perturbative corrections appearing in our series expansions, we
quote their ratio $O_{1}/O_{2}$ (at fixed point) in the second column.
The larger (absolute) values of this ratio correspond
to the better apparent convergence of truncated series.
Except for $\beta_1$ and $\gamma_{11}$, the values of $O_{1}/O_{2}$
are positive. This means that the sighns of the first- and second-order corrections
do not alter for most of the exponents.
Note, that a very similar situation has been encountered in the
analysis of perturbation expansions for surface exponents
at the ordinary transition in pure systems \cite{DS98}.

In fact, as it was shown in \cite{DS98}, the best numerical estimates 
for surface exponents in pure semi-infiinite systems have been given by 
diagonal $[1/1]$ Pad\' e approximants.
Since in the present case the qualitative behavior of underlying
series expansions is very similar, we expect to obtain the numerical results of
comparable rather good quality from nearly-diagonal two-variable
rational approximants of the types
\begin{equation}
[11/1] = \frac{1 + a_{1} u + \bar{a}_{1} v + a_{11} uv}{1 + b_{1} u +
\bar{b}_{1} v }
\end{equation}
and
\begin{equation}
[1/11] = \frac{1 + a_{1} u + \bar{a}_{1} v}{1 + b_{1} u +
\bar{b}_{1} v+ b_{11} uv }\,.
\end{equation}
The corresponding values of surface critical exponents are given in the
last two columns of the Table I.
As we can see, these numbers do not differ significantly between themselves.

The values $[0/0]$, $[1/0]$, and $[2/0]$ are simply the direct partial sums
up to the zeroth, first, and second orders, respectively.
We consider the $[11/1]$ and $[1/11]$ values as the best we could achieve 
from the available knowledge about the series expansions in the frames of 
the present approximation scheme. Their deviations from the other 
second-order estimates of the table might serve as a rough measure of the 
achieved numerical accuracy.

\section{$4-\E$ expansion}

An alternative approach to calculate the desired surface critical exponents
is the treatment of the theory in $4-\E$ space dimensions
($\sqrt\E$ expansion) and subsequent
extrapolation to $\E=1$. This approach was initiated by Ohno and Okabe \cite{OO92}
in 1992. These authors considered the two-loop approximation for correlation functions
in random semi-infinite space. They derived the corresponding series 
expansions in powers of renormalized coupling constants and $\E$, for 
surface exponents $\eta_\|$ and $\eta_\perp$. We quote their results for the 
ordinary transition in the case of present interest, setting $n=1$ in the 
expressions of Ref. \cite{OO92} and changing normalizations of coupling 
constants ($u\to v/24$, $w\to -u/3$) to fit with our notations:
\begin{eqnarray}
&&\eta_\|=2-{u\over 3}-{v\over 2}+{u^2\over 4}+{5\over 12}\,v^2+{3\over 4}\,u\,v
+\cdots\\
&&\eta_\perp=1-{u\over 6}-{v\over 4}+{5\over 36}\,u^2+{11\over 48}\,v^2+{5\over 12}\,u\,v
+\cdots\;.
\end{eqnarray}
In the present problem dots represent less important terms of order $O(\E^{3\over 2})$.
Unfortunately, only the first non-trivial corrections $\propto\!\!\sqrt\E$
have been obtained from these equations in \cite{OO92}.

Actually, it is possible to derive one more term in the $\sqrt{\E}$ expansion
of surface critical exponents using the known expressions for the 
fixed-point values up to $O(\E)$ \cite{Sh77,JK77} 
\begin{eqnarray} &&u^*=-3 \sqrt{\frac{6 \E}{53}}+18 
\frac{110+63\zeta(3)}{53^2}\,\E,\\ &&v^*=4\sqrt{\frac{6 
\E}{53}}-72\frac{19+21 \zeta(3)}{53^2}\,\E, \label{fixpe} \end{eqnarray}
where $\zeta (3)\simeq 1.2020569$ is the Riemann $\zeta$-function, and the usual
geometric factor $K_d=2^{1-d}\pi^{-d/2}/\Gamma(d/2)$ has been absorbed into 
the redefinitions of the coupling constants. Thus we obtain
\begin{eqnarray}
&& \eta_{\parallel}=2-\sqrt{\frac{6 \E}{53}}+ 
\frac{756\zeta(3)-5}{2\cdot 53^2}\, \E\,,\\
&& \eta_{\perp}=1-\frac{1}{2}\sqrt{\frac{6 \E}{53}} + 
\frac{378 \zeta(3)-29}{2\cdot 53^2}\, \E\,.\label{4e1}
\end{eqnarray}
It can be easily verified
that the above exponents satisfy the scaling relation
$\eta_{\perp} =(\eta + \eta_\parallel)/2$ with the correct value 
\begin{equation}\label{OO}
\eta=-{\E\over 106}+O(\E^{3\over 2})
\end{equation}
of the bulk theory.

Taking into account scaling relations for surface critical exponents and 
$\sqrt{\E}$ expansions for random bulk exponents $\nu$ and $\eta$ 
\cite{Sh77,JK77} we obtain, in addition, \begin{eqnarray}
&& \gamma_{11}=-\frac{1}{2}+\frac{1}{4} \sqrt{\frac{6 \E}{53}} - 
{3\over 8}\,\frac{252 \zeta(3)-37}{53^2}\;\E\,,\\ 
&& \gamma_{1}=\frac{1}{2} + 
\frac{1}{2}\sqrt{\frac{6 \E}{53}} + 
\frac{6211-1512 \zeta(3)}{8{\cdot}53^2}\; \E\,,\\ 
&& \beta_{1}=\frac{3}{4}+\frac{1}{8}\sqrt{\frac{6 
\E}{53}} - {7\over 16}\, \frac{(108 \zeta(3)-137)}{53^2}\;\E\,. 
\label{4e3} \end{eqnarray}

Similarly as in the case $d=3$ of the previous section,
we performed a Pad\' e analysis of our $\sqrt{\E}$ expansions
at $\E =1$. The numerical values of surface critical exponents
obtained in this way are represented in the Table II.

The [1/0] values for the
exponents $\eta_{\parallel}$ and $\eta_{\perp}$ reproduce the
first-order results by Ohno and Okabe \cite{OO92}.
On the other hand, the other exponents,
$\beta_{1}$, $\gamma_{11}$, and $\gamma_{1}$
slightly differ. The reason is that we calculated our $[1/0]$ estimates
directly from each respective $\sqrt{\epsilon}$ expansion,
while in Ref. \cite{OO92} they were obtained from the
scaling relations using the above numerical vales of
$\eta_{\parallel}$ and $\eta_{\perp}$.

Comparing the results from Tables I and II we see that
the values of first-order approximants, denoted by [1/0] 
and [0/1] in both cases, are of comparable magnitudes.
But, on the other hand, the values from second-order approximants
are significantly different in both tables.
The reason is that the second-order contributions of the
$\sqrt{\epsilon}$ expansion provide corrections of opposite
sighns, as compared to the "three-dimensional" theory of the
previous section (corresponding orders' ratios $O_1/O_2$
have opposite sighns).

Our choice will be in favor of our estimates derived directly in
three dimensions for the following reasons.
It is well known that to the order of $O(\E)$ the $\sqrt{\epsilon}$ expansion
fails to yield the positive correlation function exponent $\eta$
for the random bulk Ising system [see Eq. (\ref{OO})].
In fact, if we try, using the scaling relation $\eta=2\eta_\perp-\eta_\|$,
to reproduce the numerical value of $\eta$
from our second-order data of Table II, we also always obtain negative
values. This deficiency is not present in our calculations
directly at $d=3$.
Moreover, there are several reports in the literature on "bad" behavior
of the $\sqrt{\epsilon}$ expansion at larger orders, and for other
bulk exponents \cite{ShAnSo97,FHY99,FHY00}.

At the same time, the fixed-dimension massive field theory appears to give
quite regular and reliable results for random bulk systems in three dimensions
\cite{FHY99,PS00,FHY00,PV00}, even at rather low orders of perturbation theory
\cite{Jug83,HoSh92,Sh88,omega}. From our experience, the massive field theory works also well
in description of surface critical behavior in pure three-dimensional systems
\cite{DS94,DS98}.

\section{Summary}
In the present work we studied the surface critical behavior
of three-dimensional quenched random semi-infinite Ising systems
with free plane boundaries. We have calculated the corresponding surface 
critical exponents of the ordinary transition employing two alternative 
possibilities: (a) using the massive field-theoretic approach
directly in $d=3$ dimensions, and
(b) performing the $\sqrt{\epsilon}$ expansion about the upper
critical dimension $d=4$ with subsequent extrapolation to $\E=1$.
In this latter calculation we extend, to the order of $O((\sqrt{\E})^2)$,
the previous first-order results by Ohno and Okabe \cite{OO92}.

We performed a rational approximants' (Pad\'e) analysis of the 
resulting perturbation series expansions in both cases
attempting to find out the best numerical values of surface
exponents in three dimensions.
However, the typical behavior of perturbative expansions
in both calculational schemes appeared to be qualitatively different.

In the last section we gave some arguments in favor of results
obtained in the framework (a): directly at $d=3$.
Thus, the summary for our final numerical values of the surface critical
exponents at the ordinary transition in the presence of randomness is
\begin{eqnarray}
&&\eta_\|=1.36,\quad\Delta_1=0.53,\quad\eta_\perp =0.74,\quad
\beta_1 =0.88,\nonumber\\
&&\gamma_{11}=-0.24,\quad\gamma_1 =0.83,\quad
\delta_1 =2.1,\quad\delta_{11}= 0.69\,.\nonumber
\end{eqnarray}
The above values stem from the last two columns of the Table I.
The estimate of $\gamma_{11}$, for which no approximants $[11/1]$ and
$[1/11]$ exist, has been derived from the scaling relation $\gamma_{11}=\nu(1-\eta_\|)$,
where we used $\nu=0.68$ \cite{Jug83} and the above value of $\eta_\| = 1.36$.
We believe, these values are the best one can derive from all
above calculations.

The values of our exponents
characterising the surface critical behavior of semi-infinite quenched random
Ising-like systems are apparently
different from their counterparts of pure Ising systems \cite{DS94,DS98}:
$\eta_{\parallel}=1.528$, $\Delta_{1}=0.464$, $\eta_{\perp}=0.779$,
$\beta_{1}=0.796$, $\gamma_{11}=-0.333$, $\gamma_{1}=0.769$, $\delta_{1}=1.966$,
$\delta_{11}=0.582$.

We quantitatively confirm a general expectation that 
the change of the bulk universality class of a system should affect its
boundary critical behavior. So, the semi-infinite systems with quenched bulk 
disorder are characterized by a new set of the surface critical exponents. 

We suggest that the obtained results could stimulate further experimental 
work as well as numerical investigations of the boundary critical 
behavior of disordered systems.

\section*{Acknowledgments}
We should like to thank Prof. Y.\ Okabe for a useful 
discussion and Prof. H.\ W.\ Diehl for reading manuscript. M.\ Sh.\ would 
like to thank Prof. H.\ W.\ Diehl for his hospitality at the University of 
Essen. This work was supported by the National Science Council of 
the Republic of China (Taiwan) under Grant No. NSC 89-2112-M-001-084, and 
in part by the Deutsche Forschungsgemeinschaft through the Leibniz program.

\begin{table}[htb]
\caption{Surface critical exponents of the ordinary transition for $d=3$ 
up to two-loop order at the random-fixed point $
u^{*}=-0.60509,  v^{*}=2.39631$.}
\label{tab2}
\begin{center}q
\begin{tabular}{rrrrrrrrr}
\hline
$ exp $~&~
$\frac{O_{1}}{O_{2}}$~&~$[0/0]$~&~
$[1/0]$~&~$[0/1]$~&~$[2/0]$~&~$[0/2]$~&~$[11/1]$~&~$ [1/11]$\\ 
\hline
$\eta_{\parallel}$ & 2.10 & 2.00 & 1.676 & 1.721 & 1.522 & 1.581 & 
1.364 & 1.358\\

$\Delta_{1}$ & 2.35 & 0.25 & 0.412 & 0.443 & 0.481 & 0.507 & 0.530
 & 0.531\\

$\eta_{\perp}$ & 2.62 & 1.00 & 0.838 & 0.861 & 0.776 & 0.800 & 
0.736 & 0.735\\

$\beta_{1}$ & -3.11 & 0.75 & 0.912 & 0.943 & 0.860 & 0.841 & 
0.875 & 0.876\\

$\gamma_{11}$ & 0.00 & -0.50 & -0.500 & -0.500 & -0.380 & -0.364 & 
--- & ---\\

$\gamma_{1}$ & 3.71 & 0.50 & 0.743 & 0.821 & 0.808 & 
0.832 & 0.829 & 0.832\\

$\delta_{1}$ & 1.90 & 1.67 & 1.847 & 1.868 & 1.941 & 1.968 & 
2.056 & 2.054\\

$\delta_{11}$ & 1.71 & 0.33 & 0.477 & 0.501 & 0.561 & 0.595 & 
0.695 & 0.691\\
    
\end{tabular}
\end{center}
\end{table}

\begin{table}[htb]
\caption{Surface critical exponents of the ordinary transition from the 
$\sqrt{\epsilon}$expansion. } 
\label{tab3}
\begin{center}
\begin{tabular}{rrrrrrrr}
\hline
$ exp $~&~
$\frac{O_{1}}{O_{2}}$~&~$[0/0]$~&~
$[1/0]$~&~$[0/1]$~&~$[2/0]$~&~$[0/2]$~&~$[1/1]$ 
\\ 
\hline
$\eta_{\parallel}$ & -2.09 & 2.00 & 1.664 & 1.712 & 1.824 & 1.792 & 
1.772\\

$\eta_{\perp}$ & -2.22 & 1.00 & 0.832 & 0.856 & 0.907 & 0.892 & 
0.884\\

$\beta_{1}$ & 37.60 & 0.75 & 0.792 & 0.794 & 0.793 & 0.793 & 
0.793\\

$\gamma_{11}$ & -2.37 & -0.50 & -0.416 & -0.408 & -0.451 & -0.457 & 
-0.441\\

$\gamma_{1}$ & -4.17 & 0.50 & 0.668 & 0.702 & 0.628 & 
0.611 & 0.636\\
	
\end{tabular}
\end{center}
\end{table}


\begin{thebibliography}{10}
\item[$^\dag$] E-mail address: pylyp@icmp.lviv.ua
\item[$^*$]E-mail address: shpot@icmp.lviv.ua

\bibitem{Harris74}
A.~B. Harris, Journ. Phys. C {\bf 7},  1671  (1974).

\bibitem{Shalaev94}
B.~N. Shalaev, Phys. Rep. {\bf 237},  129  (1994).

\bibitem{Dots95}
V.~S. Dotsenko, Usp. Fiz. Nauk {\bf 165},  287  (1995), [{P}hys. - Usp. {\bf
  38}, 457 (1995)].

\bibitem{HL74}
A.~B. Harris and T.~C. Lubensky, Phys. Rev. Lett. {\bf 33},  1540  (1974).

\bibitem{X75}
D.~E. Khmel'nitskii, Zh. Eksp. Teor. Fiz. {\bf 68},  1960  (1975), [{S}ov.
  Phys.-JETP {\bf 41}, 981 (1975)].

\bibitem{Lub75}
T.~C. Lubensky, Phys. Rev. B {\bf 11},  3573  (1975).

\bibitem{GL76}
G. Grinstein and A. Luther, Phys. Rev. B {\bf 13},  1329  (1976).

\bibitem{B...83}
R.~J. Birgeneau {\it et~al.}, Phys. Rev. B {\bf 27},  6747  (1983).

\bibitem{M...86}
P.~W. Mitchell {\it et~al.}, Phys. Rev. B {\bf 34},  4719  (1986).

\bibitem{CS86}
D. Chowdhury and D. Stauffer, J. Stat. Phys. {\bf 44},  203  (1986).

\bibitem{MLT86}
J. Marro, A. Labarta, and J.Tejada, Phys. Rev. B {\bf 34},  347  (1986).

\bibitem{Sh77}
B. Shalaev, Zh. Eksp. Teor. Fiz. {\bf 73},  2301  (1977), [{S}ov. Phys.-JETP
  {\bf 46}, 1204 (1977)].

\bibitem{JK77}
C.Jayaprakash and H.J.Katz, Phys. Rev. B {\bf 16},  3987  (1977).

\bibitem{NR82}
K.~E. Newman and E.~K. Riedel, Phys. Rev. B {\bf 25},  264  (1982).

\bibitem{Jug83}
G. Jug, Phys. Rev. B {\bf 27},  609  (1983).

\bibitem{SSh81}
A.~I. Sokolov and B.~N. Shalaev, Fiz. Tverd. Tela {\bf 23},  2058  (1981),
  [{S}ov. Phys. Solid State {\bf 23}, 1200 (1981)]; I. O. Maier, A. I. Sokolov,
  Fiz. Tverd. Tela {\bf 26}, 3454 (1984) [{S}ov. Phys. Solid State {\bf 26},
  2076 (1984)]. The three-loop beta functions derived and used in these
  articles are in error. The correct expressions for the RG functions of the
  relevant anisotropic $mn$-componet model in three dimensions have been
  obtained, for the first time, by one of the present authors in Ref.
  \cite{Sh88}.

\bibitem{Sh88}
N.~A. Shpot, {\it On the critical behaviour of the mn-component field model in
  three dimensions: Three-loop approximation}, {K}iev ITP preprint ITF-88-140P
  (in Russian), 1988; N. A. Shpot, Phys. Lett. A {\bf 142}, 474 (1989). The
  analytical expressions for the relevant three-loop Feynman integrals in $d=3$
  are listed in Ref. \cite{BSh92}.

\bibitem{BSh92}
C. Bervillier and M. Shpot, Phys. Rev. B {\bf 46},  955  (1992).

\bibitem{ShAnSo97}
B. Shalaev, S.A.Antonenko, and A.I.Sokolov, Phys. Lett. A {\bf 230},  105
  (1997).

\bibitem{FHY99}
R. Folk, Yu. Holovatch, and T. Yavors'kii, JETP Lett. {\bf 69},  747  (1999).

\bibitem{PS00}
D.~V. Pakhnin and A.I.Sokolov, Phys. Rev. B {\bf 61},  15130  (2000).

\bibitem{FHY00}
R. Folk, Yu. Holovatch, and T. Yavors'kii, Phys. Rev. B {\bf 61},  15114
  (2000). The three-loop random bulk exponents $\nu=0.671$ and $\gamma=1.328$
  mentioned in the review part of this article have been first obtained in Ref.
  \cite{Sh88}.

\bibitem{PV00}
A. Pelissetto and E. Vicari, Phys. Rev. B {\bf 27},  6393  (2000).

\bibitem{WK98}
K.~J. Wiese and M. Kardar, Nucl. Phys. B {\bf 528},  469  (1998).

\bibitem{B-P98}
H.~G. Ballesteros {\it et~al.}, Phys. Rev. B {\bf 58},  2740  (1998).

\bibitem{WD98}
S. Wiseman and E. Domany, Phys. Rev. E {\bf 58},  2938  (1998).

\bibitem{PRR99}
G. Parisi, F. Ricci-Tersenghi, and J.~J. Ruiz-Lorenzo, Phys. Rev. E {\bf 60},
  5198  (1999).

\bibitem{Bin83}
K. Binder,  in {\em Phase Transitions and Critical Phenomena}, edited by C.
  Domb and J.~L. Lebowitz (Academic Press, London, 1983), Vol.~8, pp.\ 1--144.

\bibitem{Die86a}
H.~W. Diehl,  in {\em Phase Transitions and Critical Phenomena}, edited by C.
  Domb and J.~L. Lebowitz (Academic Press, London, 1986), Vol.~10, pp.\
  75--267.

\bibitem{Die97}
H.~W. Diehl, Int.\ J.\ Mod.\ Phys.\ B {\bf 11},  3503  (1997).

\bibitem{DN90a}
H.~W. Diehl and A. N{\"u}sser, Z.\ Phys.\ B {\bf 79},  69  (1990).

\bibitem{Die98}
H.~W. Diehl, Eur.\ Phys.\ J.\ B {\bf 1},  401  (1998).

\bibitem{PS98}
M. Pleimling and W. Selke, Eur.\ Phys.\ J.\ B {\bf 1},  385  (1998).

\bibitem{DS98}
H.~W. Diehl and M. Shpot, Nucl. Phys. B {\bf 528},  595  (1998).

\bibitem{OO92}
K. Ohno and Y. Okabe, Phys. Rev. B {\bf 46},  5917  (1992).

\bibitem{ILSS98}
F. Igl{\' o}i, P. Lajk{\' o}, W. Selke, and F. Szalma, J. Phys. A {\bf 31},
  2801  (1998).

\bibitem{LI00}
P. Lajk{\' o} and F. Igl{\' o}i, Phys. Rev. E {\bf 61},  147  (2000).

\bibitem{PCBI00}
G. Pal{\' a}gyi, C. Chatelain, B. Berche, and F. Igl{\' o}i, Eur. Phys. J. B
  {\bf 13},  357  (2000).

\bibitem{Massive}
The original approach of Parisi \cite{Par80} employing the massive field theory
  directly in fixed spatial dimensions $d<4$ has been extended to the
  semi-infinite systems with surfaces in Refs. \cite{DS94,Sh97,DS98}.

\bibitem{Par80}
G. Parisi, J. Stat. Phys. {\bf 23},  49  (1980).

\bibitem{DS94}
H.~W. Diehl and M. Shpot, Phys.\ Rev.\ Lett. {\bf 73},  3431  (1994).

\bibitem{Sh97}
M. Shpot, Cond. Mat. Phys. N 10, 143  (1997).

\bibitem{DD81c}
H.~W. Diehl and S. Dietrich, Z.\ Phys.\ B {\bf 42},  65  (1981).

\bibitem{Br59}
R. Brout, Phys. Rev. B {\bf 115},  824  (1959).

\bibitem{GL68}
R.~B. Griffiths and J.~L. Lebowitz, J. Math. Phys. {\bf 9},  1284  (1968).

\bibitem{RG81}
J. Reeve and A.~J. Guttmann, J. Phys. A {\bf 14},  3357  (1981).

\bibitem{DD80}
H.~W. Diehl and S. Dietrich, Phys.\ Lett. {\bf 80A},  408  (1980).

\bibitem{DDE83}
H.~W. Diehl, S. Dietrich, and E. Eisenriegler, Phys.\ Rev.\ B {\bf 27},  2937
  (1983).

\bibitem{Var00}
K.~B. Varnashev, Phys. Rev. B {\bf 61},  14660  (2000).

\bibitem{CPV00}
J.~M. Carmona, A. Pelissetto, and E. Vicari, Phys. Rev. B {\bf 61},  15136
  (2000).

\bibitem{Sh88a}
N. A. Shpot, Phys. Lett. A {\bf 133},  125  (1988).

\bibitem{LO}
In fact, this is an intuitive picture conveyed from the theory of bulk regular
  systems. Much less is known about the large-order behavior of perturbative
  expansions pertaining to infinite random systems (see Refs.
  \cite{lo87,McK94,AMR00}), especially at large space dimensionalities. To our
  knowledge, there are no explicit results on large orders for surface
  quantities, even in the absence of disorder.

\bibitem{lo87}
A.~J. Bray {\it et~al.}, Phys. Rev. B {\bf 36},  2212  (1987).

\bibitem{McK94}
A.~J. McKane, Phys. Rev. B {\bf 49},  12003  (1994).

\bibitem{AMR00}
G. {\' A}lvarez, V. Martin-Mayor, and J.~J. Ruiz-Lorenzo, J. Phys. A {\bf 33},
  841  (2000).

\bibitem{HoSh92}
Yu. Holovatch and M. Shpot, J. Stat. Phys. {\bf 66},  867  (1992).

\bibitem{omega}
A good example can be the correction-to-scaling exponent $\omega$. A two-loop
  result at $d=3$ is $\omega\simeq 0.45$ \cite{Jug83}. At the three-loop level,
  two different resummation schemes yield $\omega\simeq 0.359$ and $0.376$
  \cite{Sh88}. These are in an excellent agreement with the recent result
  $\omega=0.37(6)$ of high-precision Monte Carlo calculations \cite{B-P98}.

\end{thebibliography}
\end{document}